\pdfoutput=1
\documentclass[10pt]{bmc_article}

\usepackage{cite} 
\usepackage{url}  
\usepackage{ifthen}  
\usepackage{multicol}   
\usepackage[utf8]{inputenc} 
\usepackage{amssymb,amsfonts,amsmath,bbm}
\usepackage{graphicx}
\newboolean{publ}

\newenvironment{bmcformat}{\fussy\setboolean{publ}{true}}{\fussy}

\def\1{\mathbf 1}
\def\bnu{\mbox{\boldmath$\nu$}}
\def\bu{\mbox{\boldmath$u$}}
\def\brho{\mbox{\boldmath$\rho$}}

\def\vol{{\mathrm{\bf{vol}}}}

\begin{document}
\begin{bmcformat}


\title{Estimating the size of the solution space of metabolic networks}
 

\author{Alfredo Braunstein$^{1,2}$%
\email{Alfredo Braunstein - alfredo.braunstein@polito.it}%
\and
Roberto Mulet$^2$%
\email{Roberto Mulet\correspondingauthor - mulet@fisica.uh.cu}%
\and 
Andrea Pagnani$^3$%
\email{Andrea Pagnani\correspondingauthor - pagnani@isi.it}%
}

\address{%
\iid(1)ISI Foundation Viale Settimio Severo 65, Villa Gualino, I-10133 Torino, Italy\\
\iid(2)``Henri-Poincar\'e-Group'' of Complex Systems and Department of Theoretical Physics, Physics Faculty, University of Havana, La Habana, CP 10400, Cuba
\\
\iid(3)Politecnico di Torino, Corso Duca degli Abruzzi 34, I-10129, Torino, Italy}%

\maketitle


\begin{abstract}
\paragraph*{Background:}%
Cellular metabolism is one of the most investigated system of
biological interactions. While the topological nature of individual
reactions and pathways in the network is quite well understood there
is still a lack of comprehension regarding the global functional
behavior of the system. To get some insight in this direction, in the
last few years powerful theoretical methods such as extreme pathways
calculation and flux-balance analysis (FBA) have been
introduced. These methods strongly rely on the hypothesis that the
organism maximizes an objective function. However only under very
specific biological conditions ({\em e.g.}~maximization of biomass for
E-Coli in reach nutrient medium) the cell seems to obey such
optimization law. A more refined analysis not assuming extremization
remains an elusive task for large metabolic systems due to algorithmic
constraints.
\paragraph*{Results:}%
In this work we propose a novel algorithmic strategy that allows for
an efficient characterization of the whole set of stable fluxes
compatible with the metabolic constraints.  Using a technique derived
from the fields of statistical physics and information theory we
designed a stochastic algorithm to estimate the size of the affine
space containing all possible steady-state flux distributions of
metabolic networks.  The algorithm, based on the well-known Bethe
approximation, allows the computation in polynomial time of the volume
of a non full-dimensional convex polytope in high dimensions.  We show
results for toy models that compare very well with the results of
exact algorithms.  The result of our algorithm match closely the
prediction of Monte Carlo based estimations of the flux distributions
of the Red Blood Cell metabolic network. It is then used to analyze
the statistical properties of the average fluxes of the reactions in
the E-Coli metabolic network and finally to test the effect of gene
knock-outs on the size of the solution space of the E-Coli central
metabolism.
\paragraph*{Conclusions:}%
We propose a novel efficient algorithmic strategy to estimate the size
and shape of the affine space of a non full-dimensional convex
polytope in high dimensions. The method is shown to obtain,
quantitatively and qualitatively compatible results compared with
known algorithms (where this comparison is possible) being still
efficient on the analysis of large biological systems, where all other
known strategies experience an explosion in algorithmic time.

\end{abstract}

\ifthenelse{\boolean{publ}}{\begin{multicols}{2}}{}


\section*{Background}
Cellular metabolism is a complex biological problem. It can
be viewed as a chemical engine that transforms available raw materials
into energy or into the building blocks needed for the biological
function of the cells. In more specific terms a metabolic network is
indeed a processing system transforming {\em input} metabolites
(nutrients), into output metabolites (amino acids, lipids, sugars
etc.) according to very strict molecular proportions, often referred
as stoichiometric coefficients of the reactions.

Although the general topological properties of these networks are well
characterized, see for example \cite{Jeong,Fell,Zhu}, and non-trivial
pathways are well known for many species\cite{KEGG} the cooperative
role of these pathways is hard to comprehend.  In fact, the large
sizes of these networks, usually containing hundreds of metabolites
and even more reactions, makes the comprehension of the principles
that govern their global function a challenging task. Therefore, a
necessary step to achieve this goal is the use of mathematical models
and the development of novel statistical techniques to characterize
and simulate these networks.

For example, it is well known that under evolutionary pressure,
prokaryotes cells like E-Coli behave optimizing their growth
performance \cite{Ibarra}. Flux Balance Analysis (FBA) provides a
powerful tool to predict, from the whole space of phenotipic states,
which one will these cells acquire. In few words one may say that FBA
maximizes a linear function (usually the growth rate of the cell)
subject to biochemical and thermodynamic constraints
\cite{Varma}. On the other hand, cells with genetically engineered
knockouts or bacterial strains that were not exposed to evolution
pressures, need not to optimize their growth. In fact, the method of
minimization metabolic adjustment (MOMA) \cite{Segre} has shown that
knockout metabolic fluxes undergo a minimal redistribution with
respect to the flux configuration of the wild type. Yet, in more general
situations, the results are unpredictable, therefore, a tool to
characterize the shape and volume of the whole space of possible
phenotipic solutions must be welcome.

Unfortunately, this characterization has remained an elusive task. As
far as we know the attempts to obtain such a characterization were
always based on the Monte Carlo sampling of the steady-state flux
space \cite{MC}. But unfortunately, it appeared that this kind of
sampling is a very expensive calculation and is unsuitable for very
large networks.  To address this problem we propose, using a technique
derived from the fields of statistical physics and information theory,
an algorithm that may efficiently characterize the whole set of stable
fluxes compatibles with the stoichiometric constraints.

\subsection*{Mathematical Model} 
As already mentioned, a metabolic network is an engine that converts
metabolites into other metabolites through a series of intra-cellular
intermediate steps. The fundamental equation characterizing all
functional states of a reconstructed biochemical reaction network is a
mass conservation law that imposes simple linear constraints between
the incoming and outcoming fluxes at any chemical reaction. It is call
the {\em dynamic mass balance equation:}

\begin{equation}
\label{eq:stoichio}
\frac{\partial \brho}{\partial t} = 
\mathbf i  + \hat S \cdot  \bnu - \mathbf o
\end{equation}

\noindent where $\brho$ is the vector of the $M$ metabolite
concentrations in the network. $\mathbf i$ ($\mathbf o$) is the input
(output) vector of fluxes, and $\bnu$ are the reaction fluxes governed
by the $M \times N $ stoichiometric linear operator $\hat S$ (usually
named stoichiometric matrix) encoding the coefficient of the $M$
intra-cellular relations among the $N$ fluxes.

As long as just steady-state cellular properties are concerned one can
assume that a variation in the concentration of metabolites in a cell
can be ignored and considered as constant. Therefore in case of fixed
external conditions one can assume metabolites (quasi) stationarity
and consequently the {\it lhs} of \ref{eq:stoichio} can be set to
zero. Under these hypotheses the problem of finding the metabolic
fluxes compatible with flux-balance is mathematically described by the
linear system of equations

\begin{equation}
\label{eq:stoichio2}
\hat S \cdot \bnu = \mathbf o - \mathbf i \equiv \mathbf b
\end{equation}

\noindent where $\mathbf b$ is the net metabolite uptake by the cell.
Without loss of generality we can assume that the stoichiometric
matrix $\hat S$ has full rows rank, {\em i.e.}  that
$\mathrm{rank}(\hat S) = M$, since linearly dependent equations can be
easily identified and removed. Knowing that the number of metabolites
$M$ is lower than the number of fluxes $N$ the subspace of solutions
is a $(N-M)$-dimensional manifold embedded in the $N$-dimensional
space of fluxes.  In addition, the positivity of fluxes, together with
the experimentally accessible values for the maximal fluxes, limit
further the space of feasible solutions. This fact may be expressed by
the following inequalities:

\begin{equation}
\label{eq:fluxineq}
\mathbf 0 \leq \bnu \leq \bnu^{\mathrm{\bf max}}
\end{equation}

\noindent in such a way that together, \ref{eq:stoichio2} and
\ref{eq:fluxineq}, define the convex set of all the allowed
time-independent phenotipic states of a given metabolic network.

\subsection*{Sub-dimensional volumes} 
Mathematically speaking, the space of feasible solutions consistent
with the equations \ref{eq:stoichio2} constitutes an affine space
$V\subset\mathbb{R}^{N}$ of dimension $N-M$. The set of inequalities
\ref{eq:fluxineq} then defines a convex polytope $\Pi \subset V$
that, from the metabolic point of view, may be considered as the
allowed configuration space for the cell states. With the scope of
describing it, we will be interested in computing the volume of this
space and certain volumes of subspaces of it. Although conceptually
simple, the notion of sub-dimensional volume like that of $\Pi$
requires some new definitions.

Consider any linear parameterization $\phi:\mathbb{R}^{N-M}
\rightarrow V\subset\mathbb{R}^N$. A popular choice for $\phi$ is, for
instance, the inverse of the so called {\em lexicographical}
projection {\em i.e.}, the projection over the first $N-M$ coordinates
such that its restriction to $V$ has an inverse.  Being $\phi$ linear,
the $(N-M)\times N$ Jacobian matrix $\hat{\lambda}$ is constant and
coincides with the matrix of $\phi$ in the canonical bases. Denoting
$\lambda= \mathrm{det}(\hat{\lambda}^\dagger\hat{\lambda})^{\frac12}$,
the Euclidean metric in $\mathbb{R}^N$ induces a measure on $V$ (which
does not depend on $\phi$):

\begin{equation}
\label{eq:metric}
\int_V f(\bnu) d\bnu \equiv \lambda \int f(\phi(\bu)) d\bu
\end{equation}  
allowing to compute the volume of our polytope
\begin{equation}
\label{eq:vol}
\vol_V( \Pi ) \equiv 
\int_V \1_\Pi(\bnu) d\bnu = \lambda \int \1_{\phi^{-1}(\Pi)}(\bu) d\bu
\end{equation}  
\noindent where $\1_{\Pi}\left(\cdot\right)$ is the indicator
function of the set $\Pi$. It is worth pointing out that given the
linear structure of the metabolic equations, the determinant of the
mapping is a (scalar) constant.

\subsection*{Probabilistic framework}
The problem of describing the polytope $\Pi$ can formulated in a
probabilistic framework. We define the probability density $\mathcal
P$ as:
\begin{equation} 
\label{eq:polytope}
{\cal P}(\bnu ) = \vol_V(\Pi)^{-1}\1_{\Pi}(\bnu)
\end{equation}
Marginal flux probabilities over a given set of fluxes are obtained by
integrating out all remaining degrees of freedom. In particular we can
define single flux marginal probability densities as integrals on the
affine subspace $W=V\cap{\{\nu_i=\overline{\nu}\}}$:
\begin{equation}
\label{eq:single_marginal}
P_i(\nu_i ) = \int_W {\cal P}(\bnu)\prod_{j \neq i} d\nu_j  = \vol_W(\Pi\cap W)
\end{equation}
i.e. the (sub dimensional) volume of the intersection between the
polytope $\Pi$ and the hyperplane $\left\{ \nu_i =
\overline{\nu}\right\}$.


\section*{Results and Discussion}

\subsection*{Performance on low dimensional systems}

In this section we will analyze the performance of our algorithm
against an exact algorithm on low dimensional polytopes. Among the
different packages available in the Internet, we have chosen LRS
\cite{lrs1}, a program based on the {\em reverse search} algorithm
presented by Avis and Fukuda in \cite{lrs2} that can compute the
volume of non-full dimensional polytopes.  Actually, it computes the
volume of the lexicographically smallest representation of the
polytope, that for the benchmark used below, coincides with the
conventional volume estimated by our algorithm.

We have devised a specific benchmark generating random diluted
stoichiometric matrices at a given ratio $\alpha = M / N$ and fixed
number of terms different from zero $K$ in each of the reactions. All
fluxes were constrained inside the hypercube $0\leq\nu_i \leq 1$. As a
general strategy we have calculated several random instances of the
problem and measured the volume (entropy) of the polytope using the
LRS and BP algorithm. 
In particular, we have first generated 1000 realizations of random
stoichiometric matrices with $N=12,M=4$. Note that $N=12$ is around
the maximum that allows simulations with LRS in reasonable time
(around one hour per instance). For each polytope then we have
computed the two entropies $S_{LRS}$ and $S_{BP}$ with both
algorithms, fixing the same maximum value for the discretization
$q^{\mathrm{max}}=1024$ for all fluxes.

In Figure~\ref{fig:histos} we show how the quality of the BP measure
is affected by the discretization, by displaying the histogram of the
relative differences $\delta S = \frac {S_{BP} - S_{LRS}}{S_{LRS}}$
with an increasing number of bins per variable $q^{\mathrm{max}} =
16,64, 256, 1024$. One can see how a finer binning of messages
improves the quality of the approximation, seeminlgy converging to a
single distribution of errors. It is expected that for larger $N$ the
histograms would shrink: upon increasing the number of fluxes, loops
become larger and the overall topology of the graph becomes more
locally tree-like, validating the hypothesis behind the Bethe
approximation. Unfortunately, the huge increase of computer time
experimented in the calculation of the volumes using LRS made
impossible to test systems large enough to make any reasonable scaling
analysis.

Finally we address the issue of the computational complexity of the
algorithm which is a crucial one if one is interested in approaching
real world metabolic networks whose size typically is at least $50$
times the size of the largest network that can be analyzed with exact
algorithms.  In Figure \ref{fig:times} we display the running time of
both LRS and BP as a function of the number of fluxes
N. Interesting, LRS outperforms BP up to sizes $N \sim 12$ where
the running time of LRS explodes exponentially while BP maintains a
modest almost-linear behavior.

\subsection*{Distribution of fluxes in Red Blood Cell}

The algorithm was used to obtain the distribution of flux values for
each of the reactions of the Red Blood Cell metabolism.  The maximum
allowed values for the fluxes, as well as the corresponding
stoichiometric matrix were extracted directly from \cite{MC}. The
network contained 46 reactions and 34 metabolites.  Our distributions
appear in Figure \ref{fig:rbc} and as can be checked by direct
comparison they are almost identical to those obtained with the Monte
Carlo method in with the Figure 5 in~\cite{MC}. However, while the
Monte Carlo method appears to be quite expensive in computational
resources (the authors of \cite{MC} reported one week of computer
computation in a Dell Dimension 8200 to obtain their distributions)
our algorithm converged to the same results in a couple of minutes of
computation on a similar machine.

\subsection*{Analysis of gene knock-out in E-Coli}

Then, we analyze the distribution function of the average fluxes of
the metabolic network of the E-Coli. The network used contains, in its
original format, 1035 reactions and 626 metabolites,
\cite{Palssonnet}. From this network we eliminated all those reactions
with fractional stoichiometric indexes, since our algorithm in its
present form is unable to deal with them, and all those metabolites
with connectivity larger than 40. These metabolites are cofactors of
the metabolism and their elimination helps to sparse the matrix
improving the convergence of the algorithm. Finally we checked for
inconsistencies (i.e. metabolites that after the previous
transformations are only produced or consumed), and we obtain a matrix
with 1005 reactions and 560 metabolites. We ran our algorithm on this
network and computed the average fluxes in each reaction. The
probability distribution function (pdf) of these average values
appears in Figure \ref{fig:plaw_up}. As can be clearly
seen the distribution is large and may be fitted with a power law with
exponent close to 1, a result that compares very well with previous
simulations\cite{almaas,demartino} in real networks and with the
approach followed by Bianconi and Zecchina \cite{ginestra}. Moreover,
it most be mentioned that a more careful analysis of the data may
suggest that the distribution of averages fluxes has a richer
structure. In Figure \ref{fig:plaw_low} is presented
the integrated distribution function of the average fluxes of the
reactions and a clear jump appears for $\nu \approx 0.5$, and smaller
ones for $\nu \approx 0.4$ and $\mu \approx 0.6$. Whether these jumps
are just due to normal statistical fluctuations (and are correctly
smeared out in the usual binning process done to plot the pdf in
log-log scale) or reflect relevant biological or structural
information of the network is not yet known.

Finally, we concentrated our efforts in the study the E-Coli {\em
central} metabolism \cite{Palssonnet}.  To simplify the
analysis of the results, all cofactors were removed with the exception
of {\em ATP} and {\em ADP} that have been thus considered as output
and input fluxes respectively.  Under these conditions, the network
has 40 metabolites, 74 internal reactions and 10 external fluxes that
were considered irreversible following their nominal directions. The
maximum fluxes of all the reactions were considered equal to 1.

We then studied the influence of knockouts on the volume of the
solution space. In each simulation the values of the maximum flux was
kept constant and equal to 1 for all the reactions but the one with a
knockout.  For each knockout we assumed a reduction of the maximum
permitted flux $q^{\mathrm{max}}$ equal to the 0.25 of the original
value. We assumed that each reaction is knockout independently, this,
despite the fact that it is known that some reactions are associated
with the same enzyme.

The distribution of entropy change ($\Delta S=S_{KO}-S_0$) of the
solution space for the different reaction knock-outs turns out to show
two possibilities: while most of the reactions have $\Delta S> -3$
and show little and similar impact in the size of the solution space a
relevant fraction of them, has $\Delta S<-3$ but with a very uniform
distribution. To understand if there is any connection between these
reactions with large $\Delta S$ and the structure of the network, we
show in Figure \ref{fig:ds-ecoli} the changes in $S$ for the different
reaction knock-outs. In the $x-{\mathrm{axis}}$ we plot the reactions
numerated and the lines indicate the values of $\Delta S$ for each
reaction. We annotate also these reactions with $ \Delta S< -3$. Most
of them are associated with the glycolysis pathway, being more
specific, with those reactions of the glycolysis process that show
little redundancy in the topology of the network. Other reactions,
like {\em FUM, ACONT, SUC} and {\em SUCCD1i} appear in the Krebb cycle
and again show little pathway redundancy in the network.

Finally in Figure \ref{fig:correl} we display the correlations between
the changes in the entropy for different reaction knock-outs and the
most probable values $\nu^{*}$ and the average values $\bar{\nu}$ of
the fluxes in the wild network. As can be seen, two kinds of regimes
are divided by a clear threshold at $\nu \sim 0.6$ : A first one, for
small fluxes, that show consistent and slowly increasing correlations
between $\nu$ and $\Delta S$, and a second one for large fluxes, where
the correlations increase rapidly but with large fluctuations. The
presence of this threshold can be understood noting that reactions
belonging to the linear (glycolysis) an circular (Krebb cycle)
pathways are, in the wild cell, {\em fast}-flux reactions, with
average values for the fluxes larger than 0.5. Therefore once they are
knock-out, the metabolic capacities of the network are highly
affected.

\section*{Conclusions}
We proposed a novel algorithm to estimate the size and shape of the
affine space of a non full-dimensional convex polytope in high
dimensions. The algorithm was tested in specific benchmark,
i.e. random diluted stoichiometric matrices at a given ratio $\alpha =
M / N$ and fixed number of terms different from zero $K$, in each of
the reactions, with results that compare very well with those of exact
algorithms. Moreover, we show that while the running time of exact
algorithms increases more than exponentially for already moderate
sizes, our algorithm keeps a polynomial behavior for sizes as large as
$N=120$. The program was run on the Red Blood Cell metabolism, showing
with less computational effort, results that compare very well with
those previously obtained using Monte Carlo methods. Then, we
calculate the distribution of the average values of the fluxes in the
metabolism of the E-Coli and present results that are consistent with
those of the literature. Finally, our program was used to study the
E-Coli central metabolism, and we show that, as expected, reactions
with little redundancy are the ones with more impact in the size of
the space of the metabolic solutions. Specifically, most of the
reactions associated with the transformation of glucose in pyruvate,
belong to this set, as well as some reactions in the citric cycle. In
addition we show strong correlations between the characteristics of
the flux distributions of the wild type network and the changes in
size of the space of solutions after reaction knock-outs.

Let us conclude by noting that in principle the presented approach can
be extended to deal with constraints whose functional form is more
general than linear, provided that the number of variables involved in
each of the constraints remains small, as in the case of inequalities
enforcing the second low of thermodynamics for the considered
reactions \cite{Beard}. Work is in progress in this direction.

\section*{Methods}
\subsection*{Volume computation}
From a computational point of view, the problem of the exact
computation of the volume of a polytope with current methods requires
the enumeration of all its vertexes. The vertex enumeration problem is
$\# P$-hard \cite{Dyer-Frieze,Khachiyan}, but even the problem of
computing the volume, given the set of all vertexes is a big
computational challenge. Various algorithms exist for calculating the
exact volume of a polytope from its vertexes (for a review see
\cite{Buler-Enge-Fukuda}), and many software packages are available in
the Internet. Computational limitations restrict however exact
algorithmic strategies to cope with polytopes in relatively few
dimensions ({\it e.g} $N-M$ around 10 or so).  To overcome such severe
limitations we will introduce a very efficient approximate
computational strategy that will allow us to compute the volume and
the shape of the space of solutions for real-world metabolic networks.

Although the coefficient $\lambda$ could be explicitely calculated, it
turns out that as far as only relative volume quantities are
concerned, as in the case of the {\em in silico} flux knock-outs
introduced below, this term factors out and therefore we will drop it
from the rest of the computation. If one is not interested in constant
prefactors, one can proceed as follows for the integration of
Eq.~\ref{eq:polytope}. Consider the regular orthogonal grid
$\Lambda_\epsilon$ of side $\epsilon$ partitioning ${\mathbb
R}^N$. This grid maps via $\phi^{-1}$ into a partition
$\Gamma_\epsilon$ of $\phi^{-1}(\Pi)$.  The number of cells ${\mathcal
N}_\epsilon$ of $\Lambda_\epsilon$ intersecting $\Pi$ is equal to the
numbers of cells of $\Gamma_\epsilon$ intersecting
$\phi^{-1}(\Pi)$. Finally, the volume in Eq.~\ref{eq:vol} is
proportional to $\lim_{\epsilon\to 0}\epsilon^{N-M} {\mathcal
N}_\epsilon$. The same applies to the computation of the marginal of
Eq.~\ref{eq:single_marginal}, now noting that the constant
pre-factor does not depend on $\nu_i$.  The reader may wonder why one
does not directly integrate Eq.~\ref{eq:vol} by using its right hand
side, the problem is that the change of coordinates induced by $\phi$
may destruct the sparsity of the original matrix $s_{a,i}$, a
condition that turns out to be crucial for the kind of approximation
that we will use.

When dealing with integer coefficients $s_{i,a}$, as the ones
appearing in stoichiometric relations, a further simplification in the
approximate volume computation is possible: one can (always ingoring a
constant pre-factor) restrict further to integer solutions of the
system of equations. Geometrically, the space is tiled with small
hypercubes and we are actually counting the number of hypercubes
exactly fulfilling the stoichiometric equations. In summary, for any
$\epsilon$ the computation of an $\epsilon$-approximation of the
volume has been recast into a discrete combinatorial optimization
problem that can described (with a slight abuse of notation) by the
same Eqs.~\ref{eq:stoichio2} with now discrete variables $\nu_i \in
\{0,1,..., q^{\mathrm{max}}_i\}$, for $q^{\mathrm{max}}_i$ equal to
the integer part of $q^{\mathrm{max}}\times\nu_i^\mathrm{max}$, where
the integer $q^{\mathrm{max}}$ is the granularity of the
approximation.

\subsection*{Belief Propagation}
The metabolic problem can be cast into a constraint satisfaction
framework where the $M$ stoichiometric relations impose a constraint
onto a subset of the metabolic fluxes.  Let $A$ be the set of equations
and $I$ the set of fluxes. Consider the $a$-{\em th} row of $\hat S$,
and let $\{i_1,\dots,i_{n_a} \} \equiv \{i\in a \} \subset I$ be the
labels of the fluxes involved in the considered equation having
stoichiometric coefficients different from zero. Let also $\{a_1,
\cdots,a_{n_i}\}\equiv \{ a\in i \} \subset A$ be the labels of the
equations in which flux $i$ participates. The emerging structure is a
bipartite graph, with two types of nodes: {\em variable} nodes
representing the fluxes of the reactions and {\em factor} nodes
imposing mass conservation. In this case marginals become
$q$-modal probability densities that for large values of
$q^{\mathrm{max}}_i$ will approximate better and better the continuous
set of probabilities.

Under the hypothesis that the factor graph is a tree it can be shown
\cite{Baxter,Yedidia} that a given flux vector $\bnu$ satisfying all
flux-balance constraints can be expressed as a product of flux and
reaction marginals \cite{Yedidia,SumProd,BMZ}:

\begin{equation}
\label{eq:full_prob}
{\cal P}(\bnu) = \prod_{a \in A } P_{a} ( \{\nu_l\}_{l\in a} )
\prod_{i\in I} P_i ( \nu_i ) ^ {1-d_i}
\end{equation}
where $d_i$ is the number of equations in which flux $\nu_i$
participates ({\em i.e.} the degree of site $i$). The marginal
probabilities are defined as:
\begin{eqnarray}
\label{eq:marginals}
P_i(\nu_i) &=&  \sum_{\{ \nu_j  \}_{j \neq i }} {\cal P}(\bnu)\nonumber\\ 
P_{a} ( \{\nu_l\}_{l \in a} ) &=& 
\sum_{\{ \nu_j \}_{j \not \in a}}{\cal P}(\bnu)\,\,\,.
\end{eqnarray}

Belief Propagation (BP) is a local iterative algorithm that allows for
the computations of marginal probability distributions, which is exact
on trees, and perform reasonably well on locally tree-like structures
\cite{Yedidia, SumProd, BMZ, MacKay}. This approximation scheme allows
the computation of the (logarithm of the) number of solutions via
the entropy that can be expressed in terms of flux marginals:
\begin{eqnarray}
\label{eq:entropy_int}
S &\equiv& - \sum_{\bnu } {\cal P}(\bnu)\ln {\cal P}(\bnu) \nonumber \\
&=& \sum_{a \in A }\sum_{\{ \nu_j \}_{j \in a}  }
P_{a} ( \{ \nu_{j \in a} \} )
\log  P_{a} ( \{ \nu_{j \in a} \} )\nonumber \\
&-&\sum_{i\in I} 
\sum_{ \nu_i } ( d_i-1 ) P_i
( \nu_i ) \log  P_i (\nu_i)
\end{eqnarray}  
One may wonder how such an approach could be useful in a {\em
real-world} situation where the graph is not a tree. Interestingly
enough, metabolic networks are sparse, {\em i.e.} the number of
metabolites that typically participate to a certain reaction is small
with respect to the number of metabolites $M$, moreover one can
reasonably assume the typical loop length to be large enough to ensure
weak statistical dependence of neighboring sites which lay at the
heart of the Bethe approximation~\cite{PM1,PM2}. The algorithm is
based on two type of messages exchanged from variable nodes to
functional nodes, and vice versa:
\begin{itemize}
\item $\mu_{i\rightarrow a}(\nu)$: the probability that flux $i$ takes 
value $\nu$ in the absence of reaction $a$.
\item $m_{a \rightarrow i}(\nu)$: the non-normalized probability that
the balance in reaction $a$ is fulfilled given that flux $i$ takes
value $\nu$.
\end{itemize}
The two quantities satisfy the following set of functional equations:
\begin{eqnarray}
m_{a\rightarrow i} (\nu_i) &=& 
\sum_{\{ \nu_l\}_{l\in a\setminus i} }\mu_{l \rightarrow a} (\nu_l) \,\,\,
\delta\Big ( \sum_{l\in a }  s_{a,l} \nu_l \,; \, b_a \Big ) 
\nonumber \\
\mu_{i \rightarrow a } ( \nu_i ) &=& C_{i\rightarrow a} \prod_{ b \in i
  \setminus a } 
m_{b \rightarrow i}( \nu_i )
\label{eq:bp}
\end{eqnarray}
\noindent where $C_{i\rightarrow a}$ is a constant enforcing the
normalization of the probability $\mu_{i \rightarrow a }(\nu)$ and
$\delta(\cdot\, ; \cdot)$ is the Kronecker delta function.  The set of
equations \ref{eq:bp} can be solved iteratively and upon convergence
of the algorithm one can compute the marginal flux distributions as:
\begin{eqnarray}
\label{eq:marg}
P_{a} ( \{\nu_l\}_{l\in a} ) &=& 
\sum_{\{ \nu_l \}_{l \in a}} \mu_{l \rightarrow a} (\nu_l) 
\,\,\, \delta\Big ( \sum_{l\in a } s_{a,l} \nu_l\,;\, b_a \Big )
\nonumber \\ 
P_i ( \nu ) & = & \prod_{l\in i } m_{l \rightarrow i} (
\nu ) \,\,\,.
\end{eqnarray}
A brute force integration of the discrete set of equation would be
much too inefficient for analyzing large networks, due to the multiple
dimensional sum over $ \{ 0,\dots,q_l^{\mathrm{max}} \}_{l\in a
\setminus i} $ in the previous equation. A relevant speed-up in the
convergence of these equations can be achieved by noting that the
convolution product in \ref{eq:bp} can be efficiently solved with a
recursion. Let us relabel the set of fluxes in equation $a$ as $\{ l
\in a \} \equiv \{1,\dots,n_a\}$ and let assume for the sake of clarity
that $b_a=0$:
\begin{equation}
\label{eq:iteration}
 R^{(k+1)}( \nu ) =
\sum_{t=0}^{q_k^{\mathrm{max}}} R^{(k)}( \nu -
  s_{a,k} t ) \mu_{k\rightarrow a} ( t ) 
\end{equation} 
for $k \in \{0,\dots,n_a-1\}$ and initial condition $R^{(0)}( 0 ) = 1
$.  The last step of the recursion gives us $m_{a\rightarrow i} (\nu)
=R^{(n_a)}( \nu ) $. The increase of performance is substantial since
brute force integration involves a sum over $\prod_{l \in a \setminus
i } q_{l}^{\mathrm{max}}$ terms, while iteration in Eq.~ref{eq:iteration}
scales just as $\sum_{l \in a \setminus i }q_{l}^{\mathrm{max}}$.  In
some cases a faster convergence was met following a {\em refinement}
strategy in the variable $q^{\mathrm{max}}_i$. Instead of choosing a
large $q^{\mathrm{max}}_i$ from scratch and to solve the BP equations
from random initial conditions, the simulation starts using small
$q^{\mathrm{max}}_i$ that are increased after convergence. Each time
$q^{\mathrm{max}}_i$ is increased the BP equations are solved starting
from a function that fits the previous solution. With this strategy,
networks as large as 40 metabolites and 120 reactions could be
simulated in a couple minutes using a standard laptop.  It should be
noted that when the number $n_a$ is large, the computations of
$m_{a\to i}$ can be done substantially faster by means of discrete
Fourier transforms, reducing the computation time of all messages
$m_{a\to i}$ for $i\to a$ from the needed $n_a\times \sum_{l \in a
\setminus i }q_{l}^{\mathrm{max}}$ operations to just around $2\times
(\sum_{l \in a} q_{l}^{\mathrm{max}})\mathrm{log}(\sum_{l \in a}
q_{l}^{\mathrm{max}})$.

    
\section*{Authors contributions}
Authors equally contributed to this work.

\section*{Acknowledgements}
  \ifthenelse{\boolean{publ}}{\small}{}
AB was supported by Microsoft TCI grant. RM wants to thank the
International Center for Theoretical Physics in Trieste and the Center
for Molecular Immunology of La Habana, for their cordial hospitality
during the completion of this work. We are also very grateful to
Ginestra Bianconi, Michele Leone, Martin Weigt, and Riccardo Zecchina,
for very interesting discussions, and in particular to Carlotta
Martelli for sharing with us a human readable E-Coli data set.


{\ifthenelse{\boolean{publ}}{\footnotesize}{\small}
 \bibliographystyle{bmc_article}  
  \bibliography{bmc_article} }     


\ifthenelse{\boolean{publ}}{\end{multicols}}{}



\section{Figures}

\subsection{Figure~\ref{fig:times} - Running time}
\includegraphics[width=1.0\columnwidth]{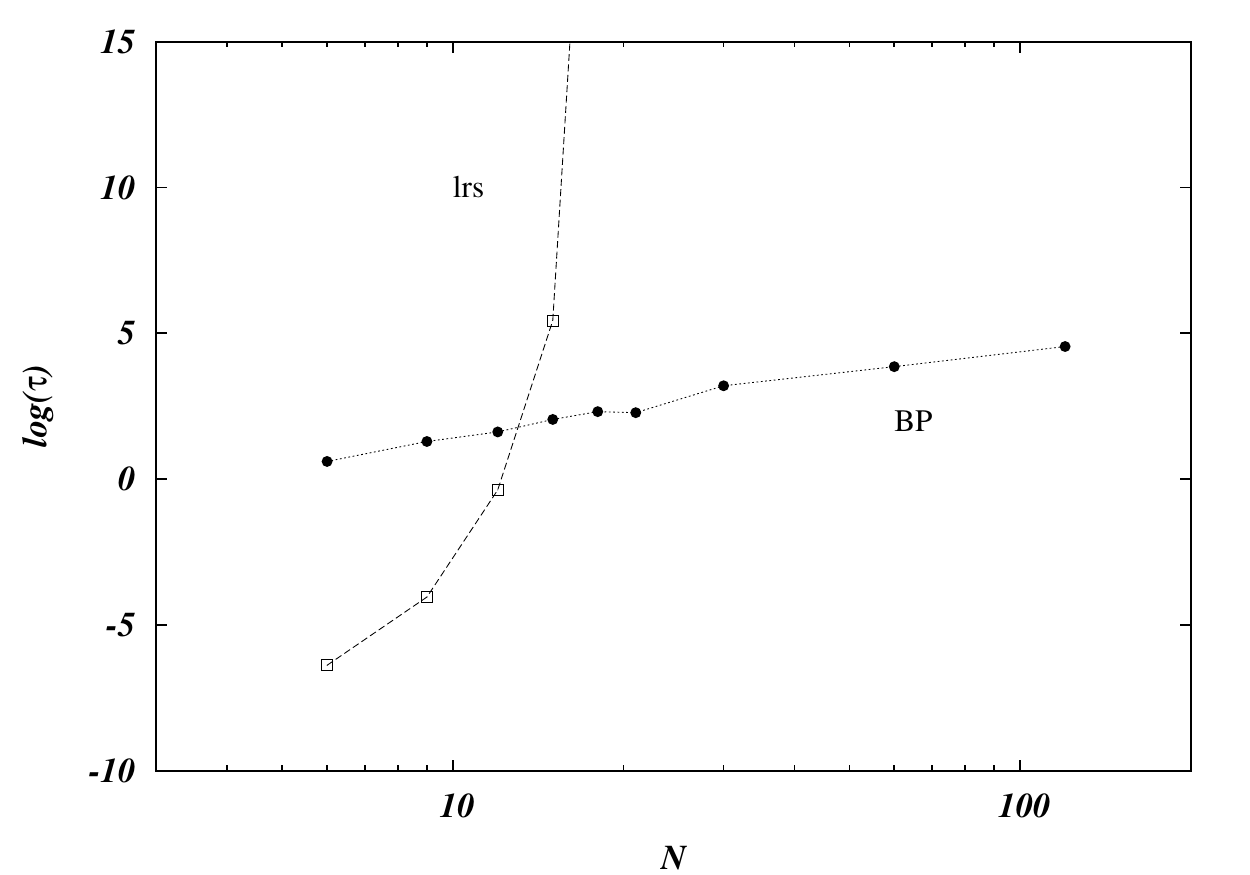}

Logarithm of the running time vs. $N$ for LRS algorithm, and
BP algorithm. Averages were taken over 5000 realizations for the
smaller lattices and 500 $N=12$.\label{fig:times}

\subsection{Figure~\ref{fig:histos} - Discrepance histogram}
\includegraphics[width=1.0\columnwidth]{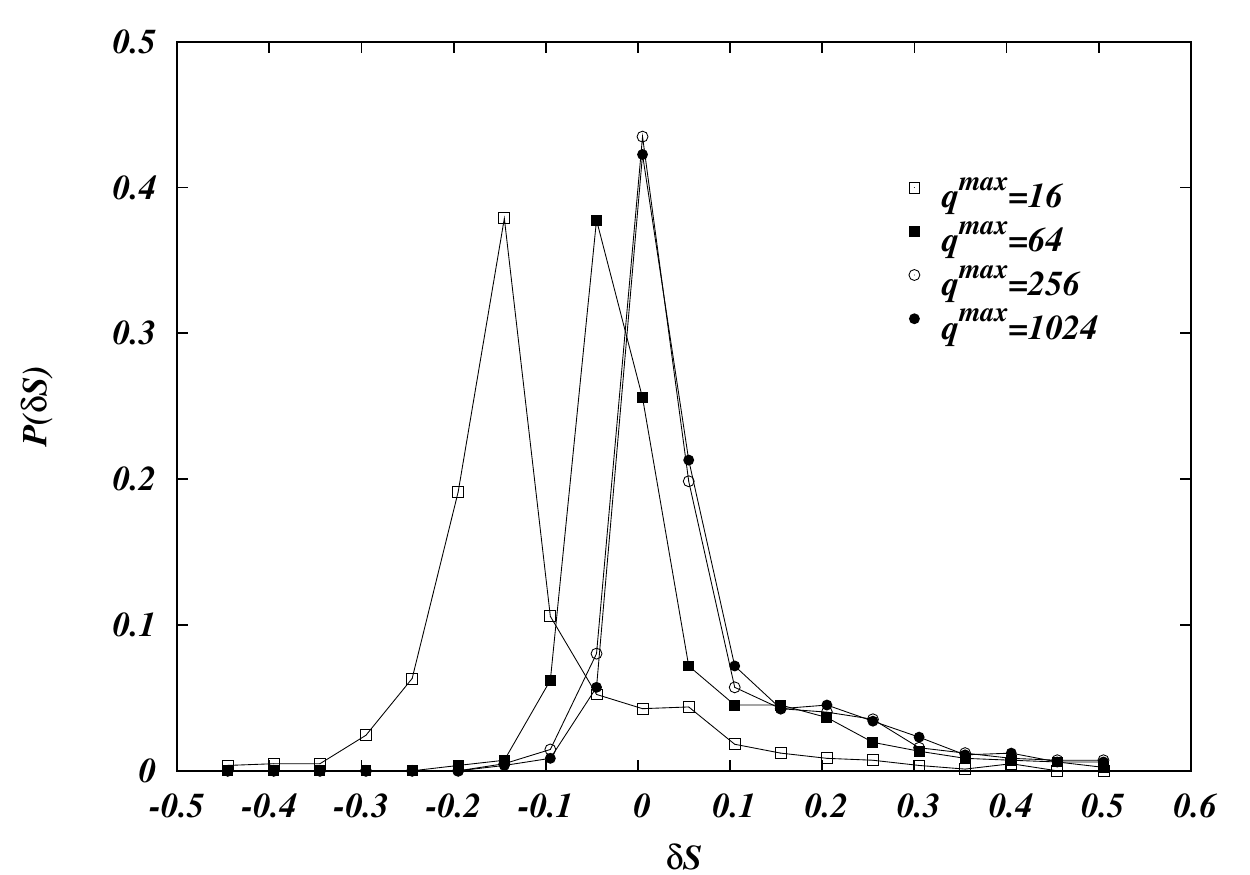}

Histograms of $\delta_S = ( S_{BP}- S_{LRS} ) / S_{LRS} $ over a set
of 1000 realizations of the stoichiometric matrix. The three
histograms are for $N=12$, $M=4$, and $K=3$, at different value of
$q^{\mathrm{max}} = 16,64, 256, 1024$.\label{fig:histos}

\subsection{Figure~\ref{fig:rbc} - Distribution of fluxes in Blood Cells}
\includegraphics[width=1.0\textwidth]{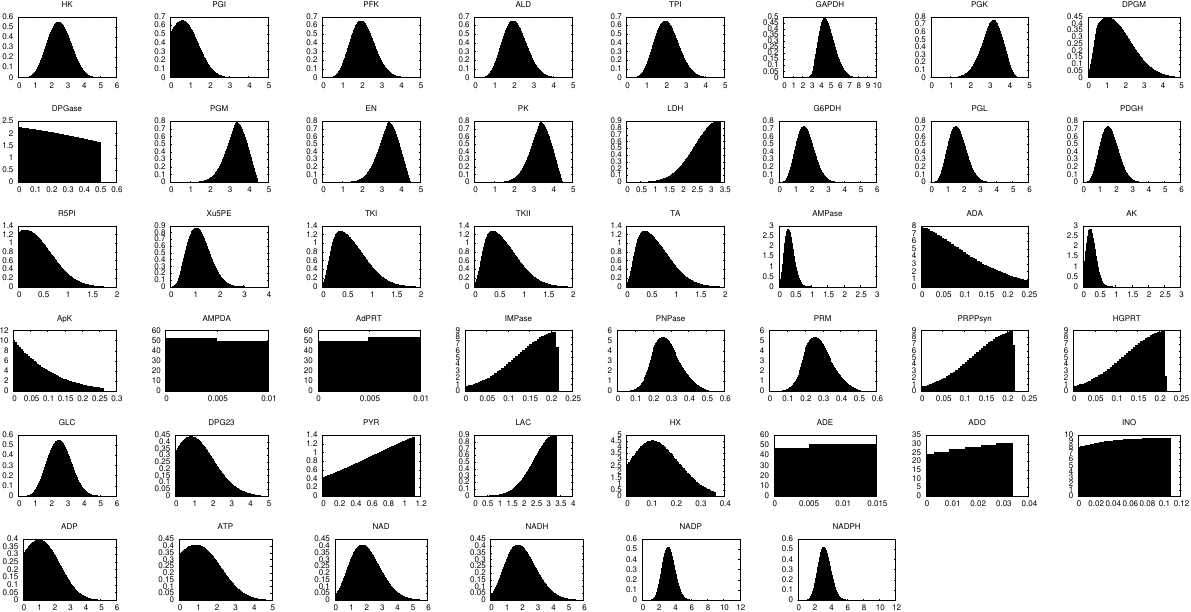}
\vspace{.5cm}

Distributions of the flux values for each reaction in the red
blood cell network. They are arranged following the same sequence as in 
\cite{MC}.
\label{fig:rbc}

\subsection{Figure~\ref{fig:plaw_up} - Distribution of average fluxes}
\includegraphics[width=0.85\columnwidth]{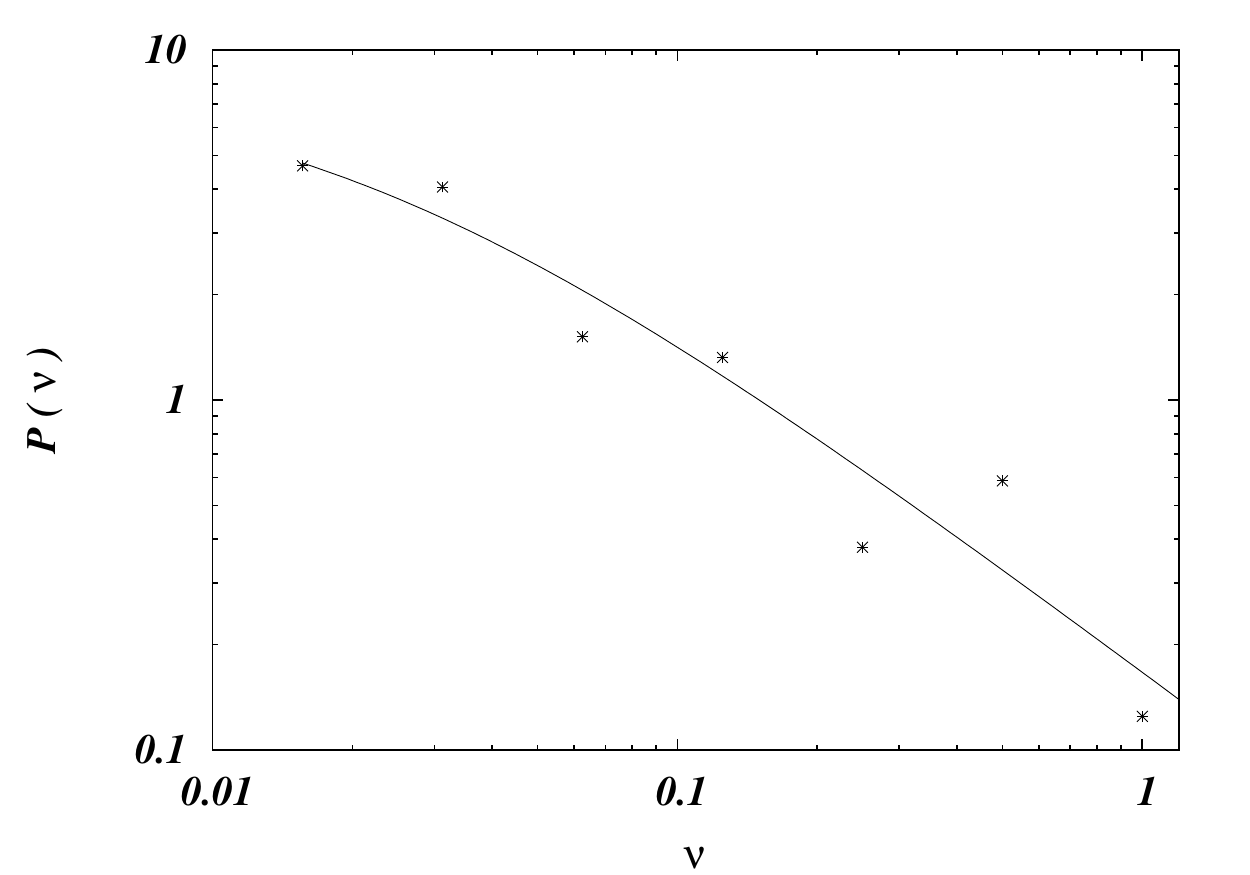}

Distribution of average fluxes ($\bar{\nu}$) of the
reactions for the E-Coli metabolism. $N=1005$ and $M=506$.\label{fig:plaw_up}

\subsection{Figure~\ref{fig:plaw_low} - Integrated distribution of average fluxes}
\includegraphics[width=0.85\columnwidth]{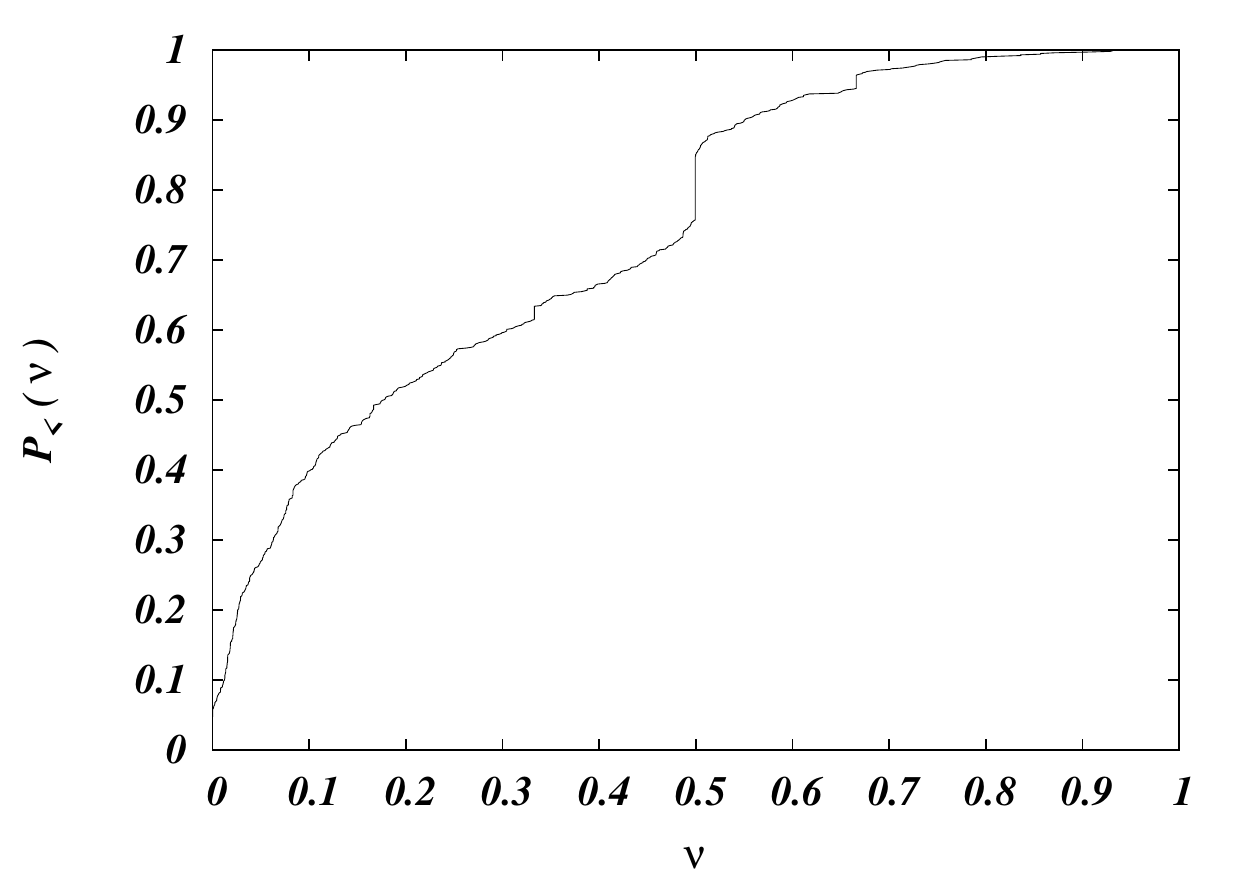} 

Integrated distribution of average fluxes ($\bar{\nu}$) of the
reactions for the E-Coli metabolism. $N=1005$ and $M=506$. Note the
jumps for $\bar{\nu}=0.4$, $\bar{\nu}=0.5$ and
$\bar{\nu}=0.6$\label{fig:plaw_low}

\subsection{Figure~\ref{fig:ds-ecoli} - Impact of knockout in E-Coli}
\includegraphics[width=0.85\columnwidth]{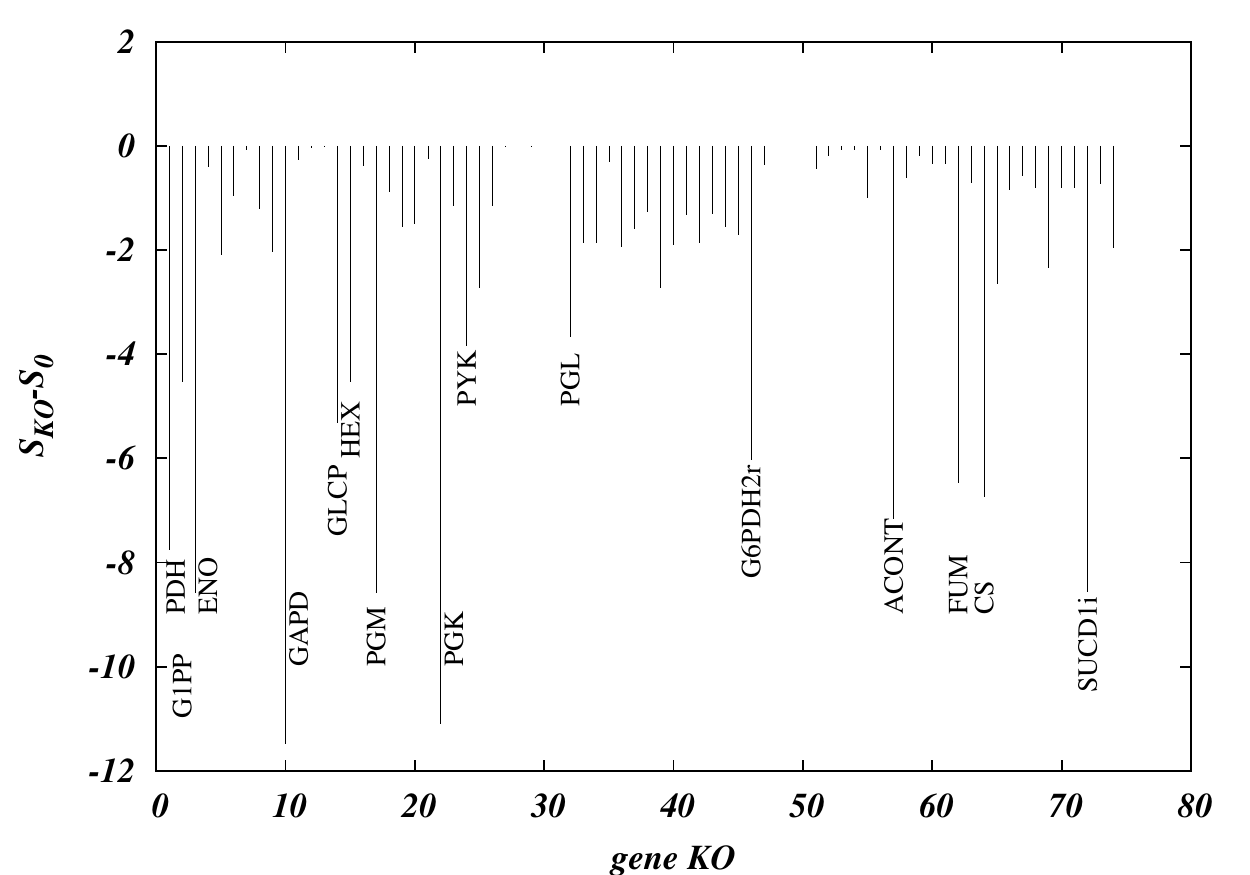} 

Changes of $S$ for different reaction knockouts in the central
metabolic network of the E-Coli. Annotated reactions belong either to
the glycolisis pathway or the Krebb cycle\label{fig:ds-ecoli}

\subsection{Figure~\ref{fig:correl} - Correlation of impact and value}
\includegraphics[width=0.85\columnwidth]{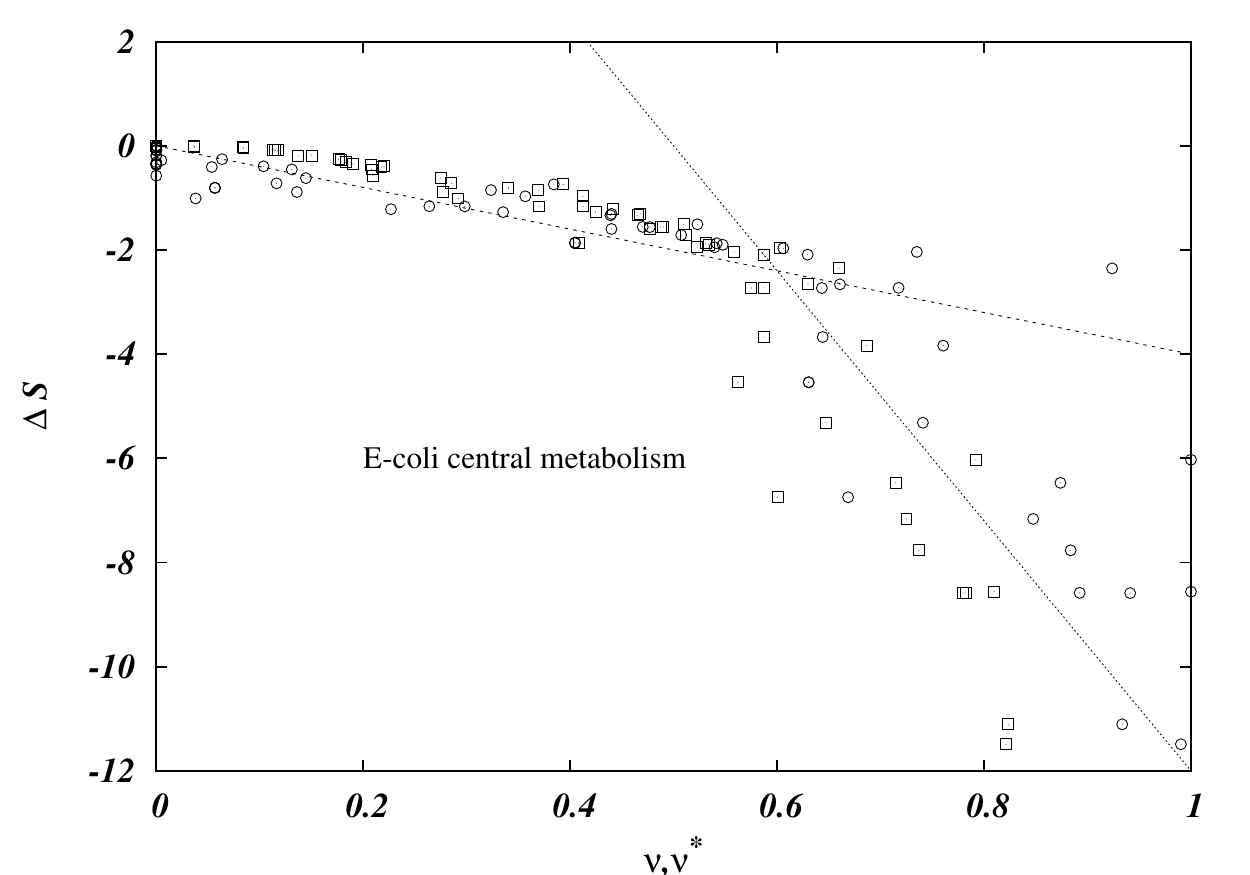}

Correlations between the change in entropy $\Delta S$ after reaction
knockouts and the average values ($\bar{\nu}$) and the most probable
values ($\nu^{*}$) of the fluxes in the central metabolism of the
E-coli. The lines are guides to the eyes\label{fig:correl}

\end{bmcformat}
\end{document}